\documentclass{amsart}
\usepackage[cp1251]{inputenc}
\usepackage{amssymb,amsmath}
\usepackage{graphicx}
\usepackage[all]{xy}
\usepackage[usenames]{color}
\usepackage{colortbl}

\begin{document}

\title[Parallel generator of $\emph{\textbf{q}}$-valued pseudorandom sequences]{Parallel generator of $\emph{\textbf{q}}$-valued pseudorandom sequences based on arithmetic polynomials}

\author{Oleg Finko}
\address[O.~Finko]{Computer Systems and Information Security of KubSTU,
Krasnodar, Moskovskaya St., 2, Russia}
\email[O.~Finko]{ofinko@yandex.ru}
\author{Dmitriy Samoylenko}
\email[D.~Samoylenko]{sam-0019@yandex.ru}
\author{Sergey Dichenko}
\email[S.~Dichenko]{dichenko.sa@yandex.ru}
\author{Nikolay Eliseev}
\email[N.~Eliseev]{n.i.eliseev@yandex.ru}

\maketitle
\begin{abstract}
                        A new method for parallel generation of $q$-valued pseudorandom sequence based on the presentation of systems generating logical formulae by means of arithmetic polynomials is proposed. Fragment consisting of $k$-elements of $q$-valued pseudorandom sequence may be obtained by means of single computing of a single recursion numerical formula. It is mentioned that the method of the ``arithmetization'' of generation may be used and further developed in order to protect the encryption gears from cryptographic onset, resulting in the initiating of mass hardware failures. The achieved results may be widely applied to the realization of perspective high-performance cryptographic facilities for information protection.
\end{abstract}
\tableofcontents \thispagestyle{empty}
\newpage

\section{Introduction}
Many specialists connect the further development of information protection facilities with the application of multiple-valued function of logical algebra (MVFLA), in particular, with the use of pseudorandom sequences (PRS) over  GF($q$) ($q>2$), which possess a wider scope of unique features, if compared with binary PRS. The most effective and approved way of obtaining PRS is the use of special switching networks called linear feedback recurrent shift register (LFSR) [1--3].

\section{General information}
  Construction of LFSR over   GF($q$) (hereinafter $q$-LFSR) is carried out by means of given characteristic polynomial:
                     $$P(z)=z^r\oplus p_{r-1}z^{r-1}\oplus p_{r-2}z^{r-2}\oplus\ \ldots\ \oplus p_{0} \pmod {q},$$
 where $P(z)\in \text{GF}
 (q)$, and  $r$ is $P(z)$ polynomial order, $r\in N$, and to the constructed according to it recurrent equation:
\begin{align}
                    s_{n+r}=p_{r-1}s_{n+r-1}\oplus p_{r-2}s_{n+r-2}\oplus\ \ldots\  \oplus p_0s_n \pmod {q},
\end{align}
$n=0, 1, 2,\ \ldots\ $;  where $p_j\in \text{GF}(q)$, $0\leq j \leq r-1$; $\oplus$~-- is the symbol of addition according to $\mod q$.

In general case $q$-LFSR consists of  $D_j $ $(j=0, 1,\ \ldots,\ r-1)$ cells and has the following initial fill: $s_0, s_1,\ \ldots,\ s_{r-1}$.
Here the ``cell'' is the  $\lceil\log_2 q \rceil$ parallel stage register ($\lceil x \rceil$  being the least integral number equal or more than $x$). After the first cycle $q$-LFSR has the following fill: $s_1, s_2,\ \ldots,\ s_{r}$. In general $q$-LFSR generates infinite $q$-valued PRS: $s_0, s_1, s_2,\ \ldots,\ s_{r-1},\ \ldots$  [2].

In  notation of linear algebra the next $q$-valued element of PRS $s_{n+r}$ is represented as a product:
 \begin{align*}
                     \begin{Vmatrix}
                                            s_{n+r} \\ s_{n+r-1} \\ \ldots\ldots\\ s_{n+2} \\ s_{n+1}  \\
                    \end{Vmatrix}^{\top}=
                    \begin{Vmatrix}
                                            s_{n+r-1} \\ s_{n+r-2}\\ \ldots\ldots \\ s_{n+1} \\ s_{n}  \\
                    \end{Vmatrix}^{\top}
                                             \cdot
                    \begin{Vmatrix}
                                            p_{r-1} &  1 &0 &  \ldots &  0 \\
                                            p_{r-2} &  0 &1 &  \ldots &  0 \\
                                             \hdotsfor[2]{5}                     \\
                                            p_1       &  0 & 0& \ldots &  1  \\
                                            p_0       &  0 &0&  \ldots &  0  \\
                    \end{Vmatrix}.
\end{align*}

In the Fig. 1 Structural diagram  of the sequential  $q$-LFSR functioning is shown.

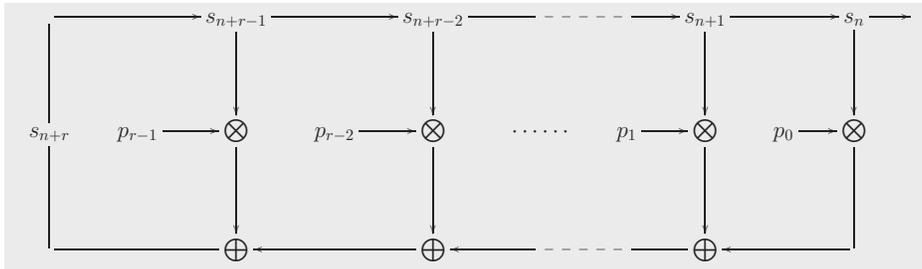
\begin{figure}[htb]
    \begin{center}\LARGE{
        \resizebox{0.99\linewidth}{!}{
\colorbox[gray]{.92}{
$$
\xymatrix{
\ar@{->}[rr]\ar@{-}[dd] &                            & s_{n+r-1}  \ar@{->}[dd] \ar@{->}[rr] &                              & s_{n+r-2}\ar@{->}[dd] \ar@{-}[r]&  \ar@{--}[r]         &   \ar@{->}[r]         &  s_{n+1}   \ar@{->}[dd] \ar@{->}[rr]     &       &  s_{n}   \ar@{->}[dd] \ar@{->}[r]  &\\
&&&&&&&&&&\\
 s_{n+r} \ar@{-}[dd]  &p_{r-1}  \ar@{->}[r] & \bigotimes \ar@{->}[dd]                      &p_{r-2} \ar@{->}[r] & \bigotimes \ar@{->}[dd]&  \cdots\cdots       &p_{1}  \ar@{->}[r]  &\bigotimes \ar@{->}[dd]  & p_{0}  \ar@{->}[r] &\bigotimes \ar@{-}[dd]                                                      & \\
 &&&&&&&&&&\\
                                      &                                &  \bigoplus \ar@{-}[ll]                          &                               & \ar@{->}[ll]\bigoplus    &   \ar@{->}[l]               &   \ar@{--}[l]                            &\ar@{-}[l] \bigoplus &                              & \ar@{->}[ll]                            &\\
 }
$$
}
}}
  \end{center}
         \begin{center}
    \caption{Structural diagram of the operation of the sequential  $q$-LFSR in accordance with formula (1)  ($\oplus$ and  $\otimes$ --- according to transaction of addition and multiplication of the $\mod q$)}
        \end{center}
\end{figure}


   As we know, PRS over GF($q$)  has a range of  ``useful'' structural properties, including [2, 3]:

 \begin{itemize}
                   \item number of symbols at the period of PRS or PRS period is defined as  $L=q^r-1$;
                   \item number of similar nonzero symbols in the PRS period is equal to  $q^{r-1}$,
                    and the number of zero symbols is equal to $q^{r-1}-1$;
                   \item addition of elements in a PRS with elements of the same PRS shifted numbering (except number equal periud repetition) gives a similar PRS shifted numbering;
                   \item autocorrelation function of PRS is defined by means of the ratio
                    $p(0)=1,$ $p(i)=-\frac{1}{q^r-1},$ $1\leq i\leq q^r-2$, etc.
 \end{itemize}

\section{Method of parallel generation of $q$-valued PRS}

In the most of practically important cases, besides the ``useful'' structural quantities, every complex system should be aimed at the achievement of some limiting characteristic or some quality indicator, what can be obtained by means of the minimization of the quantity of operations, using of resources or parallelization of computational processes of the system [4].
So, the paper [5] shows the algorithm of parallelization of generation of binary PRS based on the presentation of systems generating recurrent logical formulae by means of arithmetic polynomials. At the same time the development of computing machinery and software requires the invention of the new approaches to firmware realization both of binary functions and ($q>2$)-valued functions of logical algebra [6, 7].

Let  $s_0, s_1, s_2,\ \ldots,\ s_{r-1}, \ldots$ --
be the PRS elements, satisfying the recurrent equation (1).
  Because any element $s_n$ $(n\geq r)$ of the sequence $s_0, s_1, s_2,\ \ldots,\ s_{r-1},\ \ldots\ $
  are calculated recursively on the basis of known $r$ elements,
  let us represent the elements of PRS section $s_{n+r}, s_{n+r+1},\ \ldots,\ s_{n+2r-1}$  with the length $r$
  as the system of characteristic equations:
\begin{align}
            \begin{cases}
                                 s_{n+r}= p_{r-1}s_{n+r-1}\oplus p_{r-2}s_{n+r-2}\oplus\ \ldots\ \oplus p_0s_{n} \pmod q,  \\
                                 s_{n+r+1}= p_{r-1}s_{n+r}\oplus p_{r-2}s_{n+r-1}\oplus\ \ldots\  \oplus p_0s_{n+1} \pmod q, \\
                                 \ldots\ldots\ldots\ldots\ldots\ldots\ldots\ldots\ldots\ldots\ldots\ldots\ldots\ldots\ldots\ldots\ldots\ldots\ldots\ldots\ldots\ldots\\
                                 s_{n+2r-1}= p_{r-1}s_{n+2r-2}\oplus p_{r-2}s_{n+2r-3}\oplus\ \ldots \ \oplus p_0s_{n+r-1} \pmod q,
             \end{cases}
 \end{align}
 where $[s_{n+r}\ s_{n+r+1} \ \ldots \ s_{n+2r-1}]$  --- is the vector of PRS $r$-condition (or inner condition of $q$-LFSR on the $r$-cycle).

By the analogy with [5], let us express the right parts of the system (2) through the given initial condition:
 \begin{align}
        \begin{cases}
                    s_{n+r}=p_{r-1}s_{n+r-1}\oplus p_{r-2}s_{n+r-2}\oplus \ldots\ \oplus p_0s_n \pmod {q},  \\
                    s_{n+r+1}=p_{r-1}(p_{r-1}s_{n+r-1}\oplus p_{r-2}s_{n+r-2}\oplus \ldots \\
                    \ldots \oplus p_0s_n)\oplus p_{r-2}s_{n+r-1}\oplus \ldots  \oplus p_0s_{n+1} \pmod {q}, \\
                    \ldots\ldots\ldots\ldots\ldots\ldots\ldots\ldots\ldots\ldots\ldots\ldots\ldots\ldots\ldots\ldots\ldots\ldots\ldots\ldots\ldots\ldots\ldots\\
                    s_{n+2r-1}= p_{r-1}(p_{r-1}(p_{r-1}s_{n+r-1}\oplus p_{r-2}s_{n+r-2}\oplus \ldots\ \oplus p_0s_n)\oplus\\
                    \oplus   p_{r-2}s_{n+r-1}\oplus \ldots  \oplus p_0s_{n+1})   \oplus  p_{r-2}(p_{r-1}s_{n+r-1} \oplus p_{r-2}s_{n+r-2} \oplus \ldots\\
                     \ldots     \oplus p_0s_n)\oplus
                    \ldots   \oplus p_{0}(p_{r-1}s_{n+r-1}\oplus  p_{r-2}s_{n+r-2}\oplus  \ldots  \oplus p_0s_n)\ \ (\mathrm{mod}\ q).
        \end{cases}
 \end{align}

Let  represent the system (3) as the system $r$ MVFLA or of $r$-variables:
\begin{align}\label{4}
    \begin{cases}
     f_{1}(s_{n+r-1},  \ldots,  s_n)=p_{r-1}^{(0)}s_{n+r-1}\oplus p_{r-2}^{(0)}s_{n+r-2}\oplus \ldots \oplus p_0^{(0)}s_n \pmod q,  \smallskip  \\
 f_{2}(s_{n+r-1},    \ldots,  s_n) = p_{r-1}^{(1)}s_{n+r-1}\oplus p_{r-2}^{(1)}s_{n+r-2}\oplus\ldots  \oplus p_0^{(1)}s_n \pmod q,  \\
 \ldots\ldots\ldots\ldots\ldots\ldots\ldots\ldots\ldots\ldots\ldots\ldots\ldots\ldots\ldots\ldots\ldots\ldots\ldots\ldots\ldots\ldots\ldots\ldots\\
 f_{r}(s_{n+r-1},  \ldots,  s_n) = p_{r-1}^{(r-1)}s_{n+r-1}\oplus p_{r-2}^{(r-1)}s_{n+r-2}\oplus \ldots \oplus p_0^{(r-1)}s_{n}  \pmod q.
\end{cases}
\end{align}
Coefficients $p_{i}^{(j)}\in \{0,\ 1,\ \ldots, \ q-1\}$ are formed after reduction formulas (3).
Structural  diagram of the parallel operation of the generator in accordance with formula (4) has the form (see Fig. 2)

\begin{figure}[htb]
    \begin{center}\LARGE{
    \resizebox{0.99\linewidth}{!}{
\colorbox[gray]{.92}{
$$
\xymatrix{
\ar@{-}[r]              & s_{n} \ar@{-}[rrrrrrrrrr]        &                                      &                                   &                               &\ar@{->}[ddddddd] &                   &                                        &                                    &                                     &                            &\ar@{->}[ddddddd]  \\
\ar@{--}[rrrrrrrrrr] &                                             &                                     &                                    &\ar@{--}[ddddddd]  &                               &                   &                                         &                                   &                                      &\ar@{--}[ddddddd]&                               \\
\ar@{-}[r]             &s_{n+r-2}\ar@{-}[rrrrrrrr]     &                                     & \ar@{->}[ddd]              &                               &                              &                   &                                          &                                   &\ar@{->}[ddd]                  &                           &                               \\
\ar@{-}[r]             &s_{n+r-1}\ar@{-}[rrrrrrr]     &\ar@{->}[d]                    &                                      &                            &                              &                   &                                         &\ar@{->}[d]                  &                                       &                           &                                \\
                            &p_{r-1}^{(r-1)}  \ar@{->}[r] &\bigotimes\ar@{-}[dddd] &                                      &                          &                               &                  &p_{r-1}^{(0)} \ar@{->}[r]     &\bigotimes\ar@{-}[dddd] &                                       &                           &                               \\
                            &p_{r-2}^{(r-1)} \ar@{->}[rr] &                                    &\bigotimes\ar@{->}[dddd] &                         &                              &                    &p_{r-2}^{(0)} \ar@{->}[rr]   &                                     &\bigotimes\ar@{->}[dddd] &                            &                               \\
                            &\cdots                                  &                                     &                                     &                             &                              &                   & \cdots                                &                                    &                                       &                            &                                \\
                            &p_{0}^{(r-1)} \ar@{->}[rrrr] &                                   &                                     &                              &\bigotimes\ar@{-}[d] &                   &p_{0}^{(0)} \ar@{->}[rrrr] &                                    &                                       &                            &\bigotimes\ar@{-}[d] \\
                            &                                           &\ar@{->}[dr]                &                                     &\ar@{-->}[dl]         & \ar@{->}[dll]             &                   &                                        &\ar@{->}[dr]                 &                                       &\ar@{-->}[dl]        &\ar@{->}[dll]            \\
                            &                                          &                                     &   \bigoplus \ar@{-}[d]   &                               &                               &\cdots\cdots &                                          &                                &   \bigoplus \ar@{->}[dd]      &                            &                                \\
\ar@{=}[uuuuuuuuuu]&                                    &                                       &                  \ar@{->}[d] &                              &\ar@{=}[lllll]              &                   &\ar@{==}[ll]                     &                                 &\ar@{=}[ll]                        &                            &                                \\
                             &                                         &                                       &s_{n+2r-1}                   &                               &                               &                   &                                          &                                & s_{n+r}                           &                            &
                             }
$$
}
}}
  \end{center}
         \begin{center}
    \caption{Structural diagram of the operation of the parallel $q$-LFSR in accordance with the formula (4)}
        \end{center}
\end{figure}

   We know that the arbitrary MVFLA may be represented as arithmetical polynomial defines as [7-- 9]:
  \begin{align}\label{5}
A(S)=\sum_{i=0}^{q^{r-1}-1}a_{i}\ s_{n}^{i_0}s_{n+1}^{i_1}\ \ldots\ s_{n+r-1}^{i_{r-1}},
  \end{align}
  where $a_i$ is the $i$-ratio of arithmetical polynomial; $S=s_n,\ s_{n+1},\ \ldots,\ s_{n+r-1}$ are the arguments of MVFLA  $s_u \in {0, 1, \ldots, q-1}$ $(u=0, 1, \ldots, r-1);$
  $(i_0 \ i_1 \ \ldots \ i_{r-1})_q$ is the representation of the $i$ parameter in the $q$-ary notation system:
\begin{align*}
                                           (i_0 \ i_1 \ \ldots \ i_{r-1})_q&=\sum_{u=0}^{r-1}i_{u}q^{r-u-1}~~~~(i_{u}\in{0, 1, \ldots, q-1});\\
                    s_{u}^{i_{u}}&=
                                            \begin{cases}
                                                         1,& i_u=0,\\
                                                         s_u,& i_u\neq 0.
                                            \end{cases}
\end{align*}

 By analogy with [8] we may realize the MVFLA system (4) by calculation of some arithmetical polynomial.

 To do that, let us coordinate MVFLA (4) system with the system of arithmetical polynomials (5). Then we get:
    \begin{align}\label{6}
    \begin{cases}
     A_1(S)=\sum_{i=0}^{q^{r-1}-1}a_{1, i}\ s_{n}^{i_0}s_{n+1}^{i_1}\ \ldots\ s_{n+r-1}^{i_{r-1}},\\
     A_2(S)=\sum_{i=0}^{q^{r-1}-1}a_{2, i}\ s_{n}^{i_0}s_{n+1}^{i_1}\ \ldots\ s_{n+r-1}^{i_{r-1}},\\
      \ldots\ldots\ldots\ldots\ldots\ldots\ldots\ldots\ldots\ldots\ldots\ldots\ldots\ldots\\
       A_r(S)=\sum_{i=0}^{q^{r-1}-1}a_{r, i}\ s_{n}^{i_0}s_{n+1}^{i_1}\ \ldots\ s_{n+r-1}^{i_{r-1}}.
      \end{cases}
      \end{align}

Let  multiple the polynomials of the system (6) by weights
$q^{l-1}$   $(l=1,\, 2,\, \ldots,\, r)$:
\begin{align*}
\begin{cases}
     A_1^*(S)=q^{0}A_{1}(S)=\sum_{i=0}^{q^{r-1}-1}a_{1, i}^{*}\ s_{n}^{i_0}s_{n+1}^{i_1}\ \ldots\ s_{n+r-1}^{i_{r-1}}, \\
     A_2^*(S)=q^{1}A_{2}(S)=\sum_{i=0}^{q^{r-1}-1}a_{2, i}^{*}\ s_{n}^{i_0}s_{n+1}^{i_1}\ \ldots\ s_{n+r-1}^{i_{r-1}},\\
      \ldots\ldots\ldots\ldots\ldots\ldots\ldots\ldots\ldots\ldots\ldots\ldots\ldots\ldots\ldots\ldots\ldots\\
      A_r^*(S)=q^{r-1}A_{r}(S)\sum_{i=0}^{q^{r-1}-1}a_{r, i}^{*}\ s_{n}^{i_0}s_{n+1}^{i_1}\ \ldots\ s_{n+r-1}^{i_{r-1}},
       \end{cases}
       \end{align*}
where $a^{*}_{l,\, i}=q^{l-1}a_{l,\, i}$  $(l=1, 2, \ldots, r;$ $i=0, 1, \ldots, q^r-1).$

Then we get:
 \begin{align*}
    D(S)=\sum_{i=0}^{q^{r-1}-1}\sum_{l=1}^{d}a^{*}_{l, i}\ s_{n}^{i_0}s_{n+1}^{i_1}\ \ldots\ s_{n+r-1}^{i_{r-1}}.
       \end{align*}
 According to paper [9] the modular form of an arithmetical polynomials can be received:
\begin{align}
                        \boxed{
                    M(S)=\bigoplus_{i=0}^{q^{r-1}-1}c_{i}\ s_{n}^{i_0}s_{n+1}^{i_1}\ \ldots\ s_{n+r-1}^{i_{r-1}} \pmod {q^r},
                                }
 \end{align}
where
$$c_i=\bigoplus_{l=1}^{r}a^{*}_{l,\, i}\  \  (i=0,\, 1,\, \ldots,\, q^{r-1}-1).$$

Let  computing the values of the required MVFLA.
 To do that, we should represent the result of calculation of MVFLA in $q$-valued notation system and apply the masking operator $\Xi^t\{M(S)\}$  [9]:
  $$\Xi^t\{M(S)\}=\left \lfloor \frac{M(S)}{q^t} \right \rfloor \pmod {q},$$
  where $t$ is the required $q$-stage of the representation  $M(S)$.  Structural  diagram of the parallel operation of the generator in accordance with formula (7) has the form (see Fig. 3).

  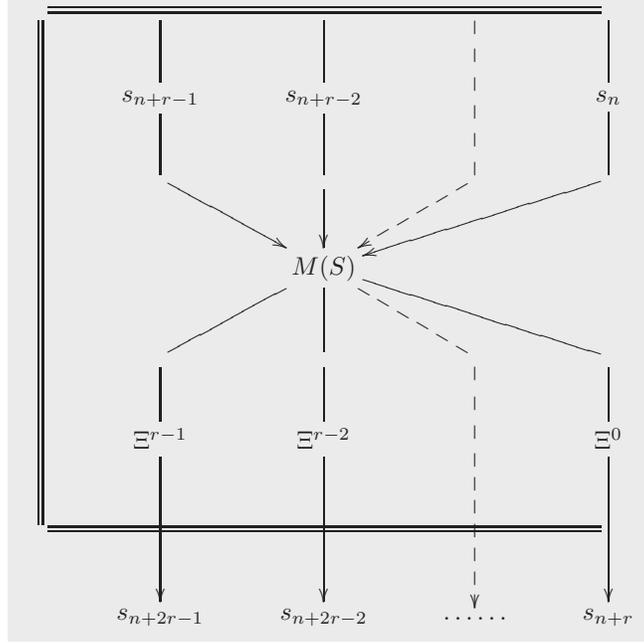
\begin{figure}[h]
    \begin{center}
    \resizebox{0.7\linewidth}{!}{
\colorbox[gray]{.92}{
$$
\xymatrix{
\ar@{=}[rrrr]                        &\ar@{-}[d]                       & \ar@{-}[d]                                                                  &\ar@{--}[d]       &\ar@{-}[d]                      \\
                                            &s_{n+r-1}  \ar@{-}[d]      &s_{n+r-2} \ar@{-}[d]                                                   & \ar@{--}[d]          & s_{n} \ar@{-}[d]           \\
                                            &\ar@{->}[dr]                   & \ar@{->}[d]                                                                & \ar@{-->}[dl]   &\ar@{->}[dll]                   \\
                                            &                                      &M(S)\ar@{-}[ld] \ar@{-}[d] \ar@{--}[dr] \ar@{-}[drr]    &                         &                                      \\
                                            &  \ar@{-}[d]                     &   \ar@{-}[d]                                                                & \ar@{--}[d]      &  \ar@{-}[d]                      \\
                                           & \Xi^{r-1} \ar@{->}[dd]     & \Xi^{r-2} \ar@{->}[dd]                                               & \ar@{-->}[dd] & \Xi^{0}  \ar@{->}[dd] \\
\ar@{=}[uuuuuu]\ar@{=}[rrrr] &                                       &                                                                                  &                        &                                       \\
                                           & s_{n+2r-1} & s_{n+2r-2} & \ldots\ldots & s_{n+r}
}
$$
}
}
  \end{center}
            \caption{Structural diagram of the operation of the parallel $q$-LFSR in accordance with the arithmetic polynomials (7)}
       \end{figure}

 \section{Numerical example}
 Let examine the construction $q=3$  LFSR, generating 3-digit PRS given by characteristic equation: $s_{k+3}=2s_{k+2}\oplus s_k \pmod {3}$ and initial fill: $s_0=0,\ s_1=1,\ s_2=2.$
The corresponding characteristic polynomial is represented as: $P(z)=z^3+2z^2+1.$

In this case the system of characteristic equations for the PRS section of three elements will be represented as follows:
\begin{align*}
                      \begin{cases}
                                        {s_{3}=2s_2\oplus s_0 \pmod {3},}  \\
                                        {s_{4}=2s_3\oplus s_1 \pmod {3},} \\
                                        {s_{5}=2s_4\oplus s_2  \pmod {3}.}
                      \end{cases}
 \end{align*}

Then let us represent the system of characteristic equations as the MVFLA system with right part of equalities, expressed by means of initial given conditions:
\begin{align*}
                      \begin{cases}
                                      f_{3}(s_2, s_1, s_0)=2s_2\oplus s_0~~({\rm mod}~3), \\
                                      f_{4}(s_2, s_1, s_0)=s_2\oplus s_1\oplus 2s_0~~({\rm mod}~3),\\
                                      f_{5}(s_2, s_1, s_0)=s_0\oplus 2s_1 \pmod {3}.
                     \end{cases}
 \end{align*}

According to (6) we shall get the system of arithmetical polynomials as follows:
\begin{align*}
                    \begin{cases}
                                         A_{3}(S)=
                                                         \frac{1}{4} (14s_2-6s_2^2+4s_0-39s_2s_0+
                                                            21s_0s_2^2+15s_0^2s_2 - 9s_0^2s_2^2), \medskip \\
                                          A_{4}(S)=\frac{1}{8} (8s_2+8s_1+42s_1s_2-30s_1s_2^2-30s_2s_1^2 +18s_1^2s_2^2+28s_0-78s_0s_2+\\
                                                            +30s_0s_2^2-78s_0s_1+78s_0s_1s_2+30s_0s_1^2-18s_0s_1^2
                                                           s_2^2-
                                                           12s_0^2
                                                           +42s_0^2s_2-\\
                                                           -18s_0^2s_2^2+42s_0^2s_1-72s_0^2s_1s_2+ 18s_0^2s_1s_2^2-18s_0^2s_1^2+
                                                           18s_0^2s_2s_1^2), \medskip \\
                                            A_{5}(S)=
                                                            \frac{1}{4}(14s_1-6s_1^2+4s_0-39s_1s_0+ 21s_0s_1^2+15s_0^2s_1- 9s_0^2s_1^2).
                      \end{cases}
 \end{align*}

Let  realize the system of arithmetical expressions as arithmetical polynomial:
\begin{align*}
\begin{cases}
& D(S)=\frac{1}{4}(14s_2-6s_2^2+4s_0-39s_2s_0+21s_0s_2^2+15s_0^2s_2- 9s_0^2s_2^2)+\\
&+3^{1}(\frac{1}{8}(8s_2+8s_1+42s_1s_2-30s_1s_2^2- 30s_2s_1^2 +18s_1^2s_2^2+28s_0-78s_0s_2+\\
&+30s_0s_2^2- 78s_0s_1+78s_0s_1s_2+30s_0s_1^2-18s_0s_1^2s_2^2-12s_0^2+42s_0^2s_2-18s_0^2s_2^2+\\
&+42s_0^2s_1-72s_0^2s_1s_2+18s_0^2s_1s_2^2-18s_0^2s_1^2+18s_0^2s_2s_1^2))+3^{2}(\frac{1}{4}(14s_1- 6s_1^2+\\
&+4s_0-39s_1s_0+21s_0s_1^2+15s_0^2s_1-9s_0^2s_1^2)).
 \end{cases}
\end{align*}

Modular polynomial form will be expressed as:

 \begin{align*}
                 & M(S)= 7s_0\oplus 9 s_0^2\oplus 21 s_1\oplus 18 s_0 s_1\oplus 9 s_0^2 s_1\oplus 18 s_0 s_1^2\oplus  20 s_2\oplus 15 s_0 s_2\oplus 6 s_0^2 s_2\oplus\\
                & \oplus 9 s_1 s_2\oplus 9 s_0 s_1 s_2\oplus  9 s_1^2 s_2\oplus 12 s_2^2\oplus 3 s_0 s_2^2\oplus 18 s_0^2 s_2^2\oplus 9 s_1 s_2^2 \pmod {27}.
 \end{align*}

According to the given initial conditions we may obtain the following three-digit fragment of PRS:
\begin{align*}
                    \mbox{step}~1
                    \begin{cases}
                    s_{3}= \Xi^{0}\{19\}=1,\\
                     s_{4}=\Xi^{1}\{19\}=0,\\
                     s_{5}=\Xi^{2}\{19\}=2;
                     \end{cases}
                                                                                                    &&
                                                                                                                                                        \mbox{step}~5
                                                                                                                                                        \begin{cases}
                                                                                                                                                         s_{15}=\Xi^{0}\{5\}=2,\\
                                                                                                                                                         s_{16}=\Xi^{1}\{5\}=1,\\
                                                                                                                                                         s_{17}=\Xi^{2}\{5\}=0;
                                                                                                                                                         \end{cases}
\\
                   \mbox{step}~2
                    \begin{cases}
                    s_{3}= \Xi^{0}\{14\}=2,\\
                     s_{4}=\Xi^{1}\{14\}=1,\\
                     s_{5}=\Xi^{2}\{14\}=1;
                     \end{cases}
                                                                                                    &&
                                                                                                                                                        \mbox{step}~6
                                                                                                                                                        \begin{cases}
                                                                                                                                                        s_{3}= \Xi^{0}\{17\}=2,\\
                                                                                                                                                         s_{4}=\Xi^{1}\{17\}=2,\\
                                                                                                                                                         s_{5}=\Xi^{2}\{17\}=1;
                                                                                                                                                         \end{cases}
\\
                    \mbox{step}~3
                     \begin{cases}
                    s_{3}= \Xi^{0}\{10\}=1,\\
                     s_{4}=\Xi^{1}\{10\}=0,\\
                     s_{5}=\Xi^{2}\{10\}=1;
                     \end{cases}
                                                                                                    &&
                                                                                                                                                        \mbox{step}~7
                                                                                                                                                        \begin{cases}
                                                                                                                                                        s_{3}= \Xi^{0}\{4\}=1,~\\
                                                                                                                                                         s_{4}=\Xi^{1}\{4\}=1,~\\
                                                                                                                                                         s_{5}=\Xi^{2}\{4\}=0;~
                                                                                                                                                         \end{cases}
\\
                    \mbox{step}~4
                     \begin{cases}
                    s_{3}= \Xi^{0}\{9\}=0,~\\
                     s_{4}=\Xi^{1}\{9\}=0,~\\
                     s_{5}=\Xi^{2}\{9\}=1;~
                     \end{cases}
                                                                                                    &&
                                                                                                                                                        \mbox{step}~8
                                                                                                                                                        \begin{cases}
                                                                                                                                                        s_{3}= \Xi^{0}\{19\}=1,\\
                                                                                                                                                         s_{4}=\Xi^{1}\{19\}=0,\\
                                                                                                                                                         s_{5}=\Xi^{2}\{19\}=0;
                                                                                                                                                         \end{cases}
\\
                                                                                                    &&
                                                                                                                                                        \cdots \cdots\cdots\cdots\cdots\cdots\cdots\cdots .
\end{align*}

\section{Conclusion}
 Here is the representation of one of the possible non-standard methods of realization of parallel algorithm of generation of $q$-valued PRS, based on the arithmetical representation of MVFLA. The developed algorithm may be used for the realization of perspective high-performance cryptographic facilities for information protection. The further direction of the research is the realization of the developed algorithm of generation of $q$-valued PRS using the redundant code redundant number system, which provide control over the errors while computing the PRS elements.

\end{document}